\documentclass[
    ,final            
  ]
  {aipproc}

\layoutstyle{6x9}
\usepackage{epstopdf}

\begin{document}

\title{Longitudinal Spin Asymmetry and Cross Section of Inclusive $\pi^0$ Production in Polarized $p+p$ Collisions at RHIC}


\author{Frank Simon (for the STAR collaboration)}{
  address={Massachusetts Institute of Technology}
}

\begin{abstract}
We present the first measurement of the cross section and the double longitudinal spin asymmetry of inclusive $\pi^0$ production in polarized p+p collisions at $\sqrt{s}$ = 200 GeV at mid-rapidity with the STAR detector, using the barrel electromagnetic calorimeter. The measured cross section is compared to NLO pQCD calculations and can provide constraints on the pion fragmentation functions. Fragmentation is studied directly by measuring the momentum fraction of $\pi^0$ in jets, a quantity that is affected by the fragmentation process and jet reconstruction effects. The double longitudinal spin asymmetry is compared to NLO pQCD calculations based on different assumptions for the gluon polarization in the nucleon to provide constraints on $\Delta g/g$. At the present level of statistics the measured asymmetry disfavors a large positive gluon polarization, but can not yet distinguish between other scenarios.

\end{abstract}

\maketitle


\section{Introduction}

One goal of the polarized p+p program at the Relativistic Heavy Ion Collider (RHIC) is the determination of the gluon polarization $\Delta G$ in the proton via spin asymmetry measurements in a variety of processes \cite{Bunce:2000uv}. Inclusive processes such as neutral pion and jet production provide sensitivity to gluons through the dominating subprocesses $gg \rightarrow gg$ and $qg \rightarrow qg$ at low and intermediate $p_t$. These probes have only modest luminosity requirements and are natural first steps in this program. The unpolarized $\pi^0$ cross section provides constraints on fragmentation functions and is an important validation of the NLO pQCD calculations used to interpret the measured spin asymmetries. Due to its large acceptance tracking and calorimetry, the STAR experiment is uniquely capable of full jet reconstruction at RHIC, thus allowing a direct study of fragmentation through the association of detected $\pi^0$ with their parent jet.

\section{Data Analysis and Results}

\begin{figure}
\includegraphics[width=0.55\textwidth]{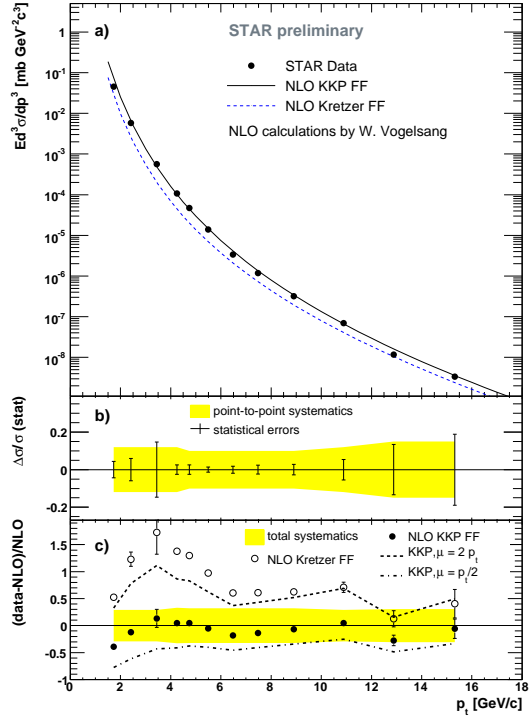}
\caption{Inclusive $\pi^0$ cross sections at mid-rapidity in p+p collisions at $\sqrt{s}$ = 200 GeV. a) compared to NLO pQCD calculations with two sets of fragmentation functions; b) relative statistical error (bars) and point-to-point systematics (band); c) relative difference to NLO pQCD calculations. For the KKP fragmentation function, the scale uncertainty is indicated by the dotted ($\mu$ = 2 p$_{\mbox{t}}$) and dashed-dotted ($\mu$ =  p$_{\mbox{t}} / 2$) lines.}
\label{fig:XSect}
\end{figure}

Neutral pions near mid-rapidity are detected with the barrel electromagnetic calorimeter BEMC \cite{Beddo:2002zx}. For the 2005 run period, half of the BEMC was fully installed and commissioned, giving a coverage of $0 < \eta < 1$ for all azimuthal angles $\varphi$. The BEMC is a lead scintillator sampling calorimeter with a granularity of $0.05 \times 0.05$ in $\Delta \eta \times \Delta \varphi$. For the reconstruction of high $p_t$ neutral pions the identification of the two decay photons with a small opening angle is crucial. This is achieved using the shower maximum detector SMD, a wire proportional counter with cathode strip readout at $\sim 5\, X_0$ depth in the calorimeter modules which provides a segmentation of $0.007 \times 0.007$ in $\Delta \eta \times \Delta \varphi$. Neutral pions are identified from the $\gamma \gamma$ invariant mass spectrum. Access to rare hard-scattering events is provided by dedicated triggers that select events with high energy deposit in one calorimeter tower, making this high-tower (HT) trigger an efficient selection tool for high $p_t$ $\pi^0$. Two such triggers, HT1 and HT2, with different thresholds were used. To achieve high detector life time for these triggers the minimum bias (MB) trigger was highly prescaled during data taking.
For the preliminary $\pi^0$ cross section presented here, a subset of the available 2005 data with an integrated luminosity of $\sim 44\, \mu\mbox{b}^{-1}$ for minimum bias and $\sim 0.4\, \mbox{pb}^{-1}$ for HT triggers was analyzed, while the preliminary asymmetry result uses HT triggers with $\sim 1.6\, \mbox{pb}^{-1}$. The cross section is given by
\vspace{-2mm}
\begin{equation}
E \frac{d^3 \sigma}{dp^3} = \frac{1}{2\pi p_t\,\Delta p_t \Delta y} \, c \, \frac{N_{\pi}}{\hat{\mathcal{L}}}
\vspace{-2mm}
\end{equation}
where $\Delta p_t$ and $\Delta y$ are the bin widths in $p_t$ and rapidity, $\hat{\mathcal{L}}$ is the sampled luminosity, $N_\pi$ is the number of reconstructed $\pi^0$ in the bin and $c$ is an overall correction factor that contains the corrections for acceptance, reconstruction and trigger efficiency thus giving the true number of pions from the number of reconstructed pions in a given bin. These corrections are determined from a full GEANT simulation of the detector, using $p+p$ events generated with the PYTHIA event generator. Good agreement between the observed and the simulated $\gamma \gamma$ spectra has been found. 

Figure \ref{fig:XSect}a) shows the preliminary cross section for inclusive $\pi^0$ production in the rapidity interval $0.1 < y < 0.9$ together with the NLO pQCD calculations \cite{Jager:2002xm} using the Kretzer \cite{Kretzer:2000yf} and the KKP \cite{Kniehl:2000hk} fragmentation function sets. Figure \ref{fig:XSect}b) shows the relative statistical errors of the data points as error bars and the preliminary point-to-point systematic errors as the shaded band. The point-to-point systematics are dominated by errors on the yield extraction from the $\gamma \gamma$ invariant mass spectra. Figure \ref{fig:XSect}c) shows the relative difference of the measured cross section to the NLO pQCD calculations for both the Kretzer and the KKP fragmentation functions. The shaded band shows the preliminary total systematic error, which is dominated by the energy scale uncertainty of the calorimeter calibration, estimated to be $\sim$5\% for the present data. The NLO calculations using the KKP fragmentation functions show excellent agreement with the data to $p_t$ below 3 GeV/c. The Kretzer fragmentation functions consistently underpredict the data. To indicate the scale uncertainty of the pQCD calculations, calculations using the KKP fragmentation functions with scales of $\mu = 2\, p_t$ and $\mu = p_t/2$ are shown by the dotted and the dashed-dotted lines, respectively. 

\begin{figure}
\begin{minipage}[t]{.5\textwidth}
  \includegraphics[width=.99\textwidth]{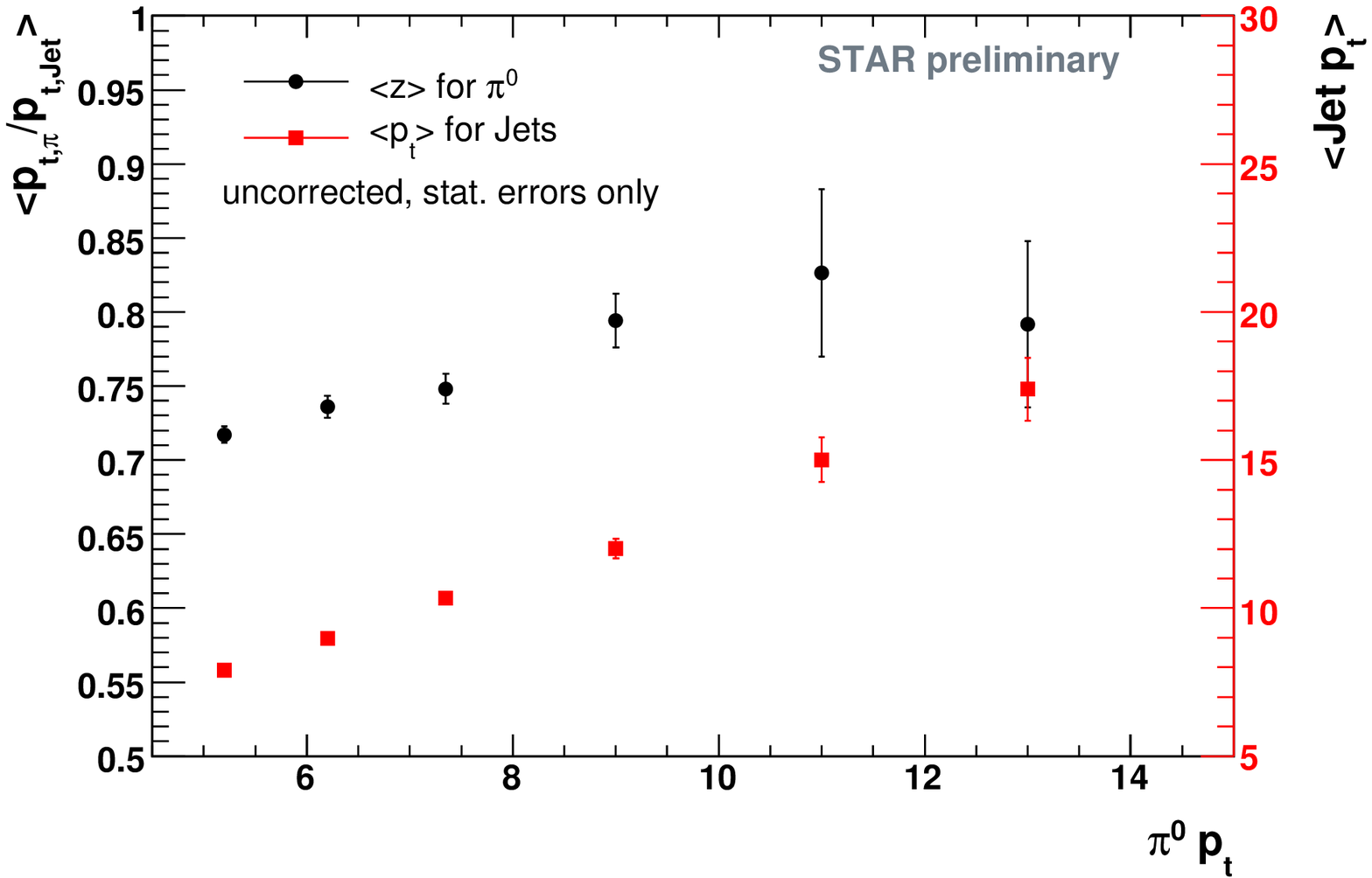}
  \end{minipage}
\begin{minipage}[t]{.5\textwidth}
  \includegraphics[width=.99\textwidth]{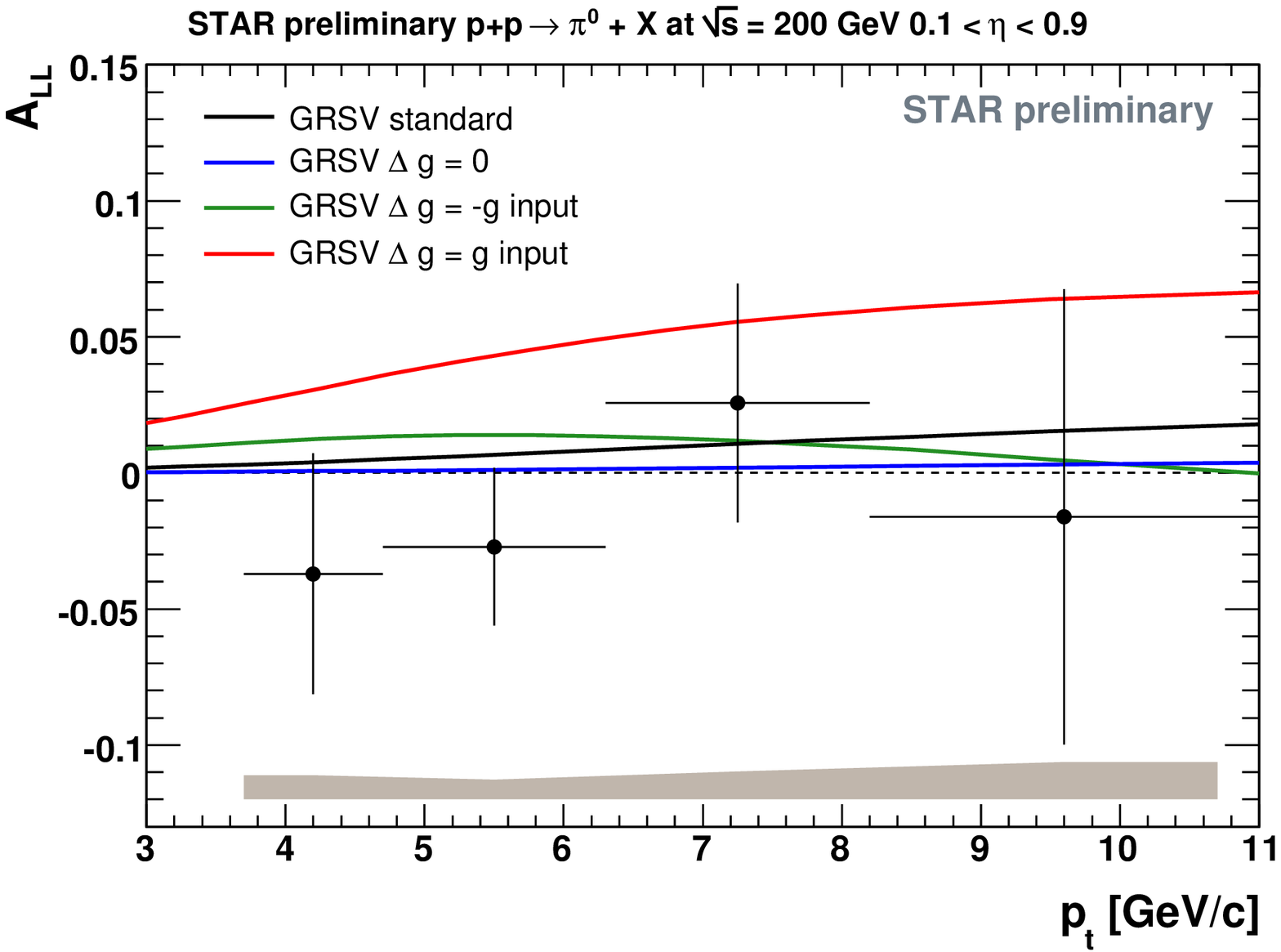}
    \label{fig:All}
 \end{minipage}
\caption{Left: Mean momentum fraction of HT1 triggered $\pi^0$ in their associated jet as a function of $p_t$ and the corresponding mean jet $p_t$. The data points are plotted at the bin center in $p_t$ and not corrected for acceptance or trigger effects. Only statistical errors are shown. \newline
Right: Double longitudinal spin asymmetry for inclusive $\pi^0$ production. The systematic error shown by the gray band does not include a 40\% normalization uncertainty due to the polarization measurement.}
  \label{fig:MeanZ} 
 \end{figure} 	

In order to investigate the momentum fraction carried by high $p_t$ $\pi^0$ in their parent jet, identified neutral pions were associated with jets found \cite{Abelev:2006uq} in the same event. An association is made if the pion is within the jet cone of 0.4 in $\eta$ and $\varphi$. To avoid edge effects, the analysis is restricted to 0.4 < $\eta$ < 0.6. To ensure the reconstructed jet is not a neutral calorimeter-only jet a maximum ratio of the neutral to total energy of 0.95 is imposed. Figure \ref{fig:MeanZ} left shows the mean momentum fraction of neutral pions associated to jets for HT1 triggers as a function of $p_t$ and the corresponding mean jet $p_t$. The momentum fraction is not corrected for acceptance, efficiency or resolution of the jet reconstruction. The mean momentum fraction of $\pi^0$ in electromagnetically triggered jets is around 0.75, and rises slightly with $p_t$, consistent with measurements of leading charged hadrons in jets in fixed-target experiments \cite{Boca:1990rh}.

The longitudinal double spin asymmetry is given by
\vspace{-2mm}
\begin{equation}
A_{LL} = \frac{1}{P_1 P_2}\frac{(N^{++} - RN^{+-})}{(N^{++} + RN^{+-})},
\vspace{-2mm}
\end{equation}
where $P_{1,2}$ are the mean measured beam polarizations and $R$ is the relative luminosity for equal and opposite beam helicities. $N^{++}$ and $N^{+-}$ are the $\pi^0$ yield in equal (++) and opposite (+-) beam helicity configurations. The polarizations are obtained with the RHIC polarimeters, typical values over the run period were $\sim$ 50\%. The relative lumionsities are monitored in STAR with the BBCs, typical $R$ values were around 1.1.

Figure \ref{fig:MeanZ} right shows the measured double spin asymmetry for $\pi^0$ production together with theoretical evaluations based on different gluon polarization scenarios \cite{Jager:2004jh}. The systematic errors shown in the figure include contributions from $\pi^0$ yield extraction and background subtraction, remaining background, possible non-longitudinal spin contributions and relative luminosity uncertainties. An overall normalization uncertainty of $\sim$ 40\% due to conservative error estimates on the preliminary polarization values is not included. Studies of parity violating single spin asymmetries and randomized spin patterns show no evidence for bunch to bunch or fill to fill systematics. The GRSV standard curve is based on the best fit to DIS data, the other curves show scenarios of extreme positive, negative and vanishing gluon polarizations. The data are consistent with three of these evaluations and tend to disfavor the scenario with a large positive gluon polarization.

\section{Summary and Outlook}

The presented inclusive neutral pion cross section in $p+p$ collisions at $\sqrt{s}$ = 200 GeV shows very good agreement with NLO pQCD calculations down to $p_t$ well below 3 GeV/c. The results favor the KKP fragmentation function set over the Kretzer set for $p_t >$ 2 GeV/c. However, scale uncertainties of the calculations are of comparable size to the difference between the fragmentation function sets. The study of the momentum fraction of $\pi^0$ in electromagnetically triggered jets show that the neutral pion carries a large part of the total jet momentum and is thus in most cases the leading particle. The preliminary double-longitudinal spin asymmetry is consistent with an evaluation based on a fit to DIS results and disfavors large positive values for the gluon polarization.



\bibliographystyle{aipproc}   
\bibliography{FSimonSpin06}

\end{document}